\documentclass[preprintnumbers,3p,12pt]{elsarticle}

\usepackage{amsmath,amssymb}
\usepackage{mathtools}
\usepackage{bm}
\usepackage{graphicx}
\usepackage{MnSymbol}
\usepackage{url}
\usepackage{hyperref}

\newcommand \bra[1]{\left< {#1} \,\right\vert}
\newcommand \ket[1]{\left\vert\, {#1} \, \right>}

\newcommand{\bea}{\begin{eqnarray}}
\newcommand{\eea}{\end{eqnarray}}
\newcommand{\simgt}{\hbox{ \raise3pt\hbox to 0pt{$>$}\raise-3pt\hbox{$\sim$} }}
\newcommand{\simlt}{\hbox{ \raise3pt\hbox to 0pt{$<$}\raise-3pt\hbox{$\sim$} }}

\newcommand{\be}{\begin{equation}}
\newcommand{\ee}{\end{equation}}

\journal{Physics Letters B}

\begin{document}

\begin{flushright}
    \normalsize TU--1307
\end{flushright}

\begin{frontmatter}

\title{Two-loop quarkonium Hamiltonian in annihilation channel}
\author{Yukinari Sumino}
 \address{Department of Physics, Tohoku University,
Sendai, 980--8578 Japan
}
\author{Takahiro Ueda}
\address{
Department of Mathematics, Juntendo University, Inzai, Chiba 270-1695, Japan
}%

\begin{abstract}
\small
%
We calculate the two-loop quarkonium Hamiltonian in the annihilation channel within the framework of potential--NRQCD effective field theory.
The result agrees with the previous calculation of the corresponding four-quark operator in NRQCD for $SU(N)$ color gauge group.
We further obtain an expression with a more general color structure applicable to other gauge groups.
Combined with the recently calculated two-loop Hamiltonian in the non-annihilation channel, this completes the full two-loop quarkonium Hamiltonian.
\end{abstract}

\end{frontmatter}


Recently, 
heavy quarkonium systems, such as charmonium, bottomonium and (would-be) toponium states, 
have become important tools for analyzing detailed aspects of the Standard Model of particle physics.
The analyses utilize the unique feature of these systems
that various properties of the QCD bound states can be calculated 
from first principles with high precision.
In fact, substantial efforts have been devoted over the years to improving the theoretical precision of heavy quarkonium observable calculations \cite{QuarkoniumWorkingGroup:2004kpm,Brambilla:2010cs}. 
Recently, progress has been marked by the computation of the two-loop heavy quarkonium Hamiltonian in the non-annihilation channel
\cite{Mishima:2024afk} as a new step towards the next-to-next-to-next-to-next-to-leading order (N$^4$LO) calculations.
The Hamiltonian is a quantum mechanical operator 
defined within the framework of the potential-NRQCD effective field theory (EFT)
\cite{Pineda:1997bj,Brambilla:2004jw}.

To complete the calculation of the quarkonium Hamiltonian at the two-loop level and up to N$^4$LO,  in this paper
we calculate the Hamiltonian in the quark-antiquark annihilation channel.
We use the calculational method developed in \cite{Mishima:2024afk}.
Namely, we match the quark-antiquark on-shell elastic scattering amplitudes between full QCD
and potential-NRQCD EFT (direct matching) at order $\alpha_s^3$ and at each order of $1/m$ expansion.
The calculation in full QCD (two-loop perturbative QCD) is performed in the following steps:
(1) Project the amplitude to a spinor basis, which projects the loop integrals to Lorentz scalar integrals.
(2) Reduce the scalar integrals into master integrals using the integration-by-parts identities.\footnote{
For the reduction of Feynman integrals,
we use \texttt{Kira}~\cite{Maierhofer:2017gsa,Klappert:2020nbg,Lange:2025fba}
interfaced to \texttt{FLINT}~\cite{flint} as well as
\texttt{Kira} interfaced to \texttt{Fermat}~\cite{fermat}, where
both results agree.
}
(3) Solve in $1/m$ expansion the differential equations satisfied by the master integrals, 
where the boundary conditions are calculated by the expansion-by-regions (EBR) method.

Explicitly, we calculate the QCD two-loop on-shell scattering amplitude for
$Q(\vec{p})+\bar{Q}(-\vec{p})= Q(\vec{p}^{\,\prime})+\bar{Q}(-\vec{p}^{\,\prime})$ in the 
color singlet channel and in the c.m.\ frame,
where 
\bea
\vec{p}^{\,\prime}=\vec{p}+\vec{k}  \,,
~~~
\left| \vec{p} \right|^2 
=\left| \vec{p}^{\,\prime}\right|^2 
\,.
\eea
The theory includes $n_h$ heavy quark flavors (each with the same mass $m$) and $n_l$ massless quark
flavors.
We calculate the amplitude corresponding to the annihilation channel of a heavy quark-antiquark pair of the same flavor.
We employ dimensional regularization, where the number of the spacetime dimensions is set to $4-2\epsilon$. 

We can utilize the calculation of the amplitude in the non-annihilation channel, for cross checks and to reduce
calculational loads, in the following way.
Apart from the color factors, Feynman diagrams for the annihilation channel are related to those of the non-annihilation channel by exchange of $s$-channel and $t$-channel.  
Hence, except for one diagram, we can obtain each amplitude (before expansion in $1/m$) from the corresponding amplitude in the non-annihilation channel.  
This applies to the expression of the amplitude in terms of the master integrals.

\begin{figure}
\begin{center}
\includegraphics[width=4cm]{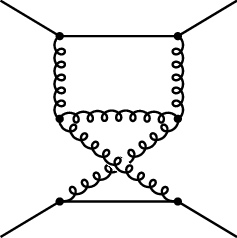}
\end{center}
\caption{\small Feynman diagram which vanishes in the non-annihilation channel after the color-singlet projection.}
\label{Fig:NonAnnDiag65}
\end{figure}

The non-annihilation-channel diagram (after projection to the color-singlet state) which cannot be converted to the annihilation-channel one
is shown in Fig.~\ref{Fig:NonAnnDiag65}.
It vanishes
due to the identity 
\bea
{\rm Tr}\left( T^a T^b T^c T^d \right) f^{ace} f^{bde} = 0
\,,
\eea
which can be shown as follows.
On the left-hand side, we rewrite $f^{ace}=-f^{cae}$ and replace $(a,b,c,d)$ by $(d,a,b,c)$.
Then it becomes the same as the original expression with the reversed sign.
On the other hand, the 
corresponding diagram in the annihilation channel does not vanish after the color-singlet projection.
Thus, apart from this diagram, we can start from evaluating each master integral in expansion in $1/m$, 
by solving the same differential equations as in the non-annihilation channel after exchanging 
the Mandelstam variables $s$ and $t$.  

\begin{figure}
\begin{center}
\includegraphics[width=6.5cm]{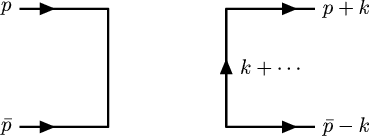}
\end{center}
\caption{\small Momentum assignment for the quark lines in the annihilation channel.
Gluon lines and loop momenta are suppressed in this diagram.
[$p=(\omega(\vec{p}),\vec{p})$, $\bar{p}=(\omega(\vec{p}),-\vec{p})$,
$k=(0,\vec{k})$ in the c.m.\ frame.]
}
\label{Fig:k-flow-in-annihilation-channel}
\end{figure}

In evaluating the master integrals in $1/m$ expansion, the contributions can be divided into those from
the HH, HS, SS, HP, SP and P-US regions from the viewpoint of the EBR method \cite{Beneke:1997zp}, where
H, S, P and US denote the hard, soft, potential and ultra-soft, respectively.
We can show that the soft contributions vanish in the annihilation channel as follows.
The only scale that can enter a soft integral is the momentum transfer $k$ in
the gluon propagators.
However, in the annihilation channel, $k$ can be chosen to flow only through the quark line, and therefore
the soft-integral is always scaleless and zero.
(See Fig.~\ref{Fig:k-flow-in-annihilation-channel}.)
Hence, there are no contributions from the HS, SS and SP regions.
The contributions from the HP and P-US regions to the scattering amplitude 
are expected to be identical with the corresponding contributions in the EFT calculation.\footnote{
This is specific to the annihilation channel, where
mixing between the soft- and potential-regions does not occur due to
absence of the soft contributions.
In the calculation of the non-annihilation channel, we need to calculate both S and P contributions
(at least naively) because there is no clear separation between the two regions.
}
Namely, they cancel in the calculation of the Hamiltonian.
Thus, we can restrict ourselves to contributions from the HH region.  
Besides the two-loop Hamiltonian, there is no contribution  from the EFT to the amplitude,
corresponding to the HH region of full QCD.
Hence, the HH contribution of full QCD is
directly matched to the two-loop Hamiltonian of the EFT in the annihilation channel.

Contributions from the HH region to the master integrals
can be expressed by 10 two-loop master integrals of the HH region.
Expansions of these HH master integrals in $\epsilon$ are given in \cite{Gerlach:2019kfo}.
We have checked these expressions, partly analytically, and all of them numerically
by using \texttt{pySecDec} \cite{Borowka:2017idc,Borowka:2018goh,Heinrich:2021dbf,Heinrich:2023til}.

Let us explain our convention of the spinor basis.
Originally we choose a spinor basis in terms of the $\gamma$ matrices in dimensional regularization. 
It can be easily expressed by a two-component
spinor basis using the Pauli matrices in dimensional regularization,
satisfying
\bea
\{ \sigma^i, \sigma^j \}=2\delta^{ij}\mathbb{I}\,,
~~\delta^{ii}=3-2\epsilon\,,
~~{\rm tr}\,\mathbb{I}=2\,.
\label{PauliArithmetic}
\eea
Eventually
the two-loop Hamiltonian in the annihilation channel up to N$^4$LO can be expressed with 3 spinor structures,
\bea
&&
\bar{\Lambda}^{\rm An}_1=\sigma^a \otimes \sigma^a\,,
 \nonumber \\ &&
\bar{\Lambda}^{\rm An}_2=\sigma^a\,\sigma^b\, \sigma^c \otimes \sigma^c\,\sigma^b\, \sigma^a
\,,
\label{Eq:ann-spinor-basis}
 \\ &&
\bar{\Lambda}^{\rm An}_3=\sigma^a\,\sigma^b\,\sigma^c\,\sigma^d\,\sigma^e\, \otimes \sigma^e\,\sigma^d\,\sigma^c\,\sigma^b\,\sigma^a
 \,.
\nonumber
\eea
By way of example, contraction with the initial ($\xi,\zeta$) and final ($\xi',\zeta'$) spinors are defined as
\bea
\bra{\xi',\zeta'}\bar{\Lambda}^{\rm An}_2\ket{\xi,\zeta}=
\left(\xi'^\dagger \sigma^a\,\sigma^b\, \sigma^c\, \tilde{\zeta}' \right) 
\left(\tilde{\zeta}^\dagger \sigma^c\,\sigma^b\, \sigma^a\, \xi \right) \,, ~~\text{etc.},
\eea
reflecting the spinor structure of the color current in the annihilation channel.
Here, $\xi^{(\prime)}$ and $\zeta^{(\prime)}$ denote, respectively, the two-component
spinor wave functions for the quark and antiquark, which are related to the four-component 
Dirac spinor wave functions\footnote{
The normalization of the spinor wave function is chosen to match the normalization of
the scattering
amplitude in
non-relativistic quantum mechanics.
}
as
\bea
u(\vec{p})=\sqrt{\textstyle \frac{\omega(\vec{p})+m}{2\omega(\vec{p})}}
\left(\begin{array}{c}
\xi\\
\frac{\vec{p}\cdot\vec{\sigma}}{\omega(\vec{p})+m}\xi
\end{array}\right) ,
~~~
v(\vec{p})=\sqrt{\textstyle \frac{\omega(\vec{p})+m}{2\omega(\vec{p})}}
\left(\begin{array}{c}
\frac{\vec{p}\cdot\vec{\sigma}}{\omega(\vec{p})+m}\tilde{\zeta}\\
\tilde{\zeta}
\end{array}\right) ,
\eea
with\footnote{
$\xi,\zeta$ belong to the $\boldsymbol{2}$ representation of the spin $SU(2)$ group, whereas
$\tilde{\zeta}$ belongs to the $\boldsymbol{2}^*$ representation.
}
\bea
\tilde{\zeta}=-i\sigma_2 \zeta^*
\,,
~~~~~~
\omega(\vec{p})=\sqrt{\vec{p}^2+m^2}
\,.
\eea
We assume existence of the charge conjugation matrix in dimensional regularization with the property
\bea
(i\sigma_2)(-i\sigma_2)=\mathbb{I}
\,,
~~~
(i\sigma_2)\sigma^a(-i\sigma_2)=-{\sigma^a}^T
\,.
\eea

The Hamiltonian in the annihilation channel before expansion in $\epsilon$ is given in the form
\bea
H_{\rm An}=\frac{C_F \,\bar{\mu}^{-2\epsilon}}{ k^2} \sum_{i=1}^3 \sum_{j=2}^3
\sum_{n,\ell\ge 0}\left(g_R^2 \,\bar{\mu}^{2\epsilon}\right)^j W^{\rm An}_{\{i,j,n,2\ell\}} \left(\frac{k}{m}\right)^n 
\frac{(p^2)^\ell+(p'^2)^\ell}{2\,m^{2\ell}}\, \bar{\Lambda}^{\rm An}_i\,,
\label{defW}
\eea
where 
\bea
k=\left| \vec{k} \right| \,,
~~~
p^2=\left| \vec{p} \right|^2 \,,
~~~
p^{\prime\,2}=\left| \vec{p}^{\,\prime}\right|^2 \,,
~~~
\vec{p}^{\,\prime}=\vec{p}+\vec{k}  \,.
\eea
$g_R=\sqrt{4\pi \alpha_s^{(n_h+n_l)}(\mu)}$ denotes the renormalized gauge coupling constant in the $\overline{\rm MS}$ scheme of the full theory (with $n_h$ heavy quark flavors and
$n_l$ massless quark flavors);
$\bar{\mu}^2=\mu^2 \,e^{\gamma_E}/(4\pi) $, where $\gamma_E=0.5772\dots$ denotes the Euler constant.

There is no contribution at tree level since the intermediate state cannot be a single gluon which is color octet.
At one loop  and  up to ${\cal O}(\alpha_s\beta^3)$ relative to LO, non-zero Wilson coefficients are given by
\bea
&&
W^{\rm An}_{\{1,2,2,0\}}=
m^{-2\epsilon} \left(-\frac{(\epsilon-1) (6 \epsilon-7) iI_H}{4 (2
   \epsilon-3)}-\frac{(6 \epsilon-7) iI_H^{\rm An}}{4 (2 \epsilon-3)}\right)
\,,
 \\ && 
W^{\rm An}_{\{2,2,2,0\}}=
m^{-2\epsilon} 
    \left(-\frac{(\epsilon-1) iI_H}{4 (2
   \epsilon-3)}-\frac{iI_H^{\rm An}}{4 (2 \epsilon-3)}\right)
   \,.
\eea
The one-loop master integrals of the hard region in the EBR
method are given by
\bea
&&
i I_H=
(4 \pi )^{\epsilon -2} \Gamma (\epsilon -1)
\,,
~~~~~~
iI_H^{\rm An}=
- \frac{2^{2 \epsilon -5} e^{i \pi  \epsilon } \pi ^{\epsilon -\frac{1}{2}} \csc (\pi  \epsilon
   )}{\Gamma \left(\frac{3}{2}-\epsilon \right)}
\,, 
~~~
\eea
(after factoring out the dimensionful parameters).

The list of two-loop Wilson coefficients is lengthy.
We provide the corresponding non-zero $W^{\rm An}_{\{i,3,n,2\ell\}}$ as a list in
a separate file \cite{MathematicaFile} readable, e.g., by {\it Mathematica}.
The counter term and two-loop master integrals included in the list are given in the Appendix.

Next we expand the coefficients
of $\bar{\Lambda}^{\rm An}_i$ in $\epsilon$
while we ignore any ${\cal O}(\epsilon)$ contributions in $\bar{\Lambda}^{\rm An}_i$.
(Namely, we simply take the limit $\epsilon= 0$ for $\bar{\Lambda}^{\rm An}_i$.)
The Hamiltonian is given in the form\footnote{
It is most natural to express the Hamiltonian in terms of $\alpha_s(m)$ since
only the hard region contributes.
Here, however, we adjust the expression to match that of the non-annihilation channel \cite{Mishima:2024afk}.
}
\bea
H_{\rm An}=\frac{16\pi^2 C_F }{ k^2} \sum_{i=1}^4 \sum_{j=2}^3
\sum_{n,\ell\ge 0}\left(\frac{\alpha_s(k)}{4\pi}\right)^j C^{\rm An}_{\{i,j,n,2\ell\}} \left(\frac{k}{m}\right)^n 
\frac{(p^2)^\ell+(p'^2)^\ell}{2\,m^{2\ell}}
\, O_i\,.
\label{Eq: HamiltonianAn}
\eea
$\alpha_s(k)$ denotes the strong coupling constant in the $\overline{\rm MS}$ scheme of the theory with
$n_l$ flavors only, renormalized at $\mu=k$.\footnote{
It is customary to express the Hamiltonian in terms of
the coupling constant of 
the theory with
$n_l$ flavors only $\alpha_s(\mu)\equiv \alpha_s^{(n_l)}(\mu)$.
}
The spinor basis in three dimensions is defined as
\bea
&&
O_1=\mathbb{I}\otimes\mathbb{I}\,,
~~
O_2=\vec{S}^2 \,,
~~
O_3=\frac{i}{k^2}\,\vec{S}\cdot\left(\vec{p}\times\vec{k}\right)\,,
~~
O_4=\sigma^a \otimes \sigma^a-\frac{3}{k^2}\left(\vec{k}\cdot\vec{\sigma}\right)\otimes\left(\vec{k}\cdot\vec{\sigma}\right)\,,
~~~
\nonumber\\
\eea
with 
\bea
\vec{S}=\frac{\vec{\sigma}}{2}\otimes \mathbb{I} + \mathbb{I}\otimes\frac{\vec{\sigma}}{2} \,.
\eea
We have performed Fierz transformation and converted the spinor basis to the standard one
in quantum mechanics.
Namely, contraction with the spinors are defined as
\bea
\bra{\xi',\zeta'}\sigma^a \otimes \sigma^a\ket{\xi,\zeta}=
\left(\xi'^\dagger \sigma^a\, \xi \right) 
\left({\zeta'}^\dagger \sigma^a\, \zeta \right) \,, ~~\text{etc.}
\eea

The one-loop Wilson coefficients are well known and finite \cite{Kniehl:2002br}.
Non-zero ones up to N$^4$LO are given by
\bea
&&
C^{\rm An}_{\{1,2,2,0\} }= -2-i \pi +2 \log (2)
\,,
\\&&
C^{\rm An}_{\{2,2,2,0\} }= 1+\frac{i \pi }{2}-\log (2)
\,.
\eea
The two-loop Wilson coefficients are also finite.
Non-zero ones up to N$^4$LO are given by
\bea
\begin{aligned}
C^{\rm An}_{\{1,3,2,0\}} =\;&
C_A \Bigg[
  \frac{44}{3} \log(2)\log\!\left(\frac{k^2}{m^2}\right)
  - \frac{22}{3} i\pi \log\!\left(\frac{k^2}{m^2}\right)
  - \frac{44}{3} \log\!\left(\frac{k^2}{m^2}\right)
\\ &\quad
  - \frac{39}{2}\zeta(3)
  + \frac{13}{6} i\pi^3
  - \frac{5\pi^2}{9}
  - \frac{398}{9} i\pi
  - \frac{589}{9}
\\ &\quad
  - \frac{44}{3}\log^2(2)
  + \frac{26}{3}\pi^2 \log(2)
  + \frac{44}{3} i\pi \log(2)
  + \frac{796}{9}\log(2)
\Bigg]
\\[4pt]
&+ C_F \Bigg[
  21\zeta(3) + 92 + 20 i\pi
  - \frac{16\pi^2}{3} - i\pi^3
  - 40\log(2) - 4\pi^2 \log(2)
\Bigg]
\\[4pt]
&+ n_l \Bigg[
  - \frac{8}{3} \log(2)\log\!\left(\frac{k^2}{m^2}\right)
  + \frac{4}{3} i\pi \log\!\left(\frac{k^2}{m^2}\right)
  + \frac{8}{3} \log\!\left(\frac{k^2}{m^2}\right)
\\ &\quad
  - \frac{10\pi^2}{9}
  + \frac{32}{9} i\pi
  + \frac{64}{9}
  + \frac{8}{3}\log^2(2)
  - \frac{8}{3} i\pi \log(2)
  - \frac{64}{9}\log(2)
\Bigg]
\\[4pt]
&
  - \frac{2\pi^2}{9}n_h
\,,
\end{aligned}
\eea
\bea
\begin{aligned}
C^{\rm An}_{\{2,3,2,0\}} =\;&
C_A \Bigg[
  - \frac{22}{3} \log(2)\log\!\left(\frac{k^2}{m^2}\right)
  + \frac{11}{3} i\pi \log\!\left(\frac{k^2}{m^2}\right)
  + \frac{22}{3} \log\!\left(\frac{k^2}{m^2}\right)
\\ &\quad
  + \frac{39}{4}\zeta(3)
  - \frac{13}{12} i\pi^3
  + \frac{5\pi^2}{18}
  + \frac{199}{9} i\pi
  + \frac{589}{18}
\\ &\quad
  + \frac{22}{3}\log^2(2)
  - \frac{13}{3}\pi^2 \log(2)
  - \frac{22}{3} i\pi \log(2)
  - \frac{398}{9}\log(2)
\Bigg]
\\[4pt]
&+ C_F \Bigg[
  - \frac{21}{2}\zeta(3) - 46 - 10 i\pi
  + \frac{8\pi^2}{3}
  + \frac{i\pi^3}{2}
  + 20\log(2)
  + 2\pi^2 \log(2)
\Bigg]
\\[4pt]
&+ \frac{d_F^{abc}d_F^{abc}}{{N}\,C_F}
\Bigg[
  \frac{3}{8}\zeta(3) + \frac{3}{4} + i\pi
  - \frac{\pi^2}{72}
  - \frac{i\pi^3}{9}
  - 2\log(2)
  - \frac{1}{36}\pi^2 \log(2)
\Bigg]
\\[4pt]
&+ n_l \Bigg[
  \frac{4}{3} \log(2)\log\!\left(\frac{k^2}{m^2}\right)
  - \frac{2}{3} i\pi \log\!\left(\frac{k^2}{m^2}\right)
  - \frac{4}{3} \log\!\left(\frac{k^2}{m^2}\right)
\\ &\quad
  + \frac{5\pi^2}{9}
  - \frac{16}{9} i\pi
  - \frac{32}{9}
  - \frac{4}{3}\log^2(2)
  + \frac{4}{3} i\pi \log(2)
  + \frac{32}{9}\log(2)
\Bigg]
\\[4pt]
&
+ \frac{\pi^2}{9}n_h
\,,
\end{aligned}
\eea
where
$\{ T_F^a, T_F^b \}=d_F^{abc}T_F^c$ and $N={\rm tr}_C(\mathbf{1})$. 
In QCD [$SU(3)$ gauge group], the values of the color factors read
$C_A=3$, $C_F=4/3$, and $d_F^{abc}d_F^{abc}/N=40/9$.

\begin{figure}
\begin{center}
\includegraphics[width=4.5cm]{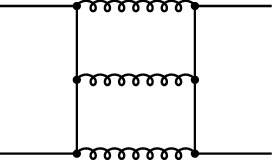}
\end{center}
\caption{\small Example diagram including the color factor $d_F^{abc}d_F^{abc}/N$ after the color-singlet projection.}
\label{Fig:Diag-3gluonInterMed}
\end{figure}

In the $SU(N)$ case,  after appropriate conversion of the spinor basis,\footnote{
The relation between our spinor basis and that of \cite{Gerlach:2019kfo} is given by
\bea
\Sigma_0^{c,(1)}\otimes\Sigma_0^{c,(2)}=\bar{\Lambda}_1^{\rm An},
~~~
\Sigma_1^{c,(1)}\otimes\Sigma_1^{c,(2)}=
\biggl(-\frac{11}{2}+5\epsilon\biggr)\bar{\Lambda}_1^{\rm An}
+\frac{1}{2}\bar{\Lambda}_2^{\rm An}
,
\nonumber\\
\Sigma_2^{c,(1)}\otimes\Sigma_2^{c,(2)}=
\biggl(\frac{65}{4}-71\epsilon+49\epsilon^2\biggr)\bar{\Lambda}_1^{\rm An}
+\biggl(-\frac{7}{2}+9\epsilon\biggr)\bar{\Lambda}_2^{\rm An}
+\frac{1}{4}\bar{\Lambda}_3^{\rm An}
\,,
\eea
where the commutators in \cite{Gerlach:2019kfo} are rewritten as $[A,B]=AB-BA$ and 
$\sigma^a$'s in each term are reordered
using the anitcommutators $\{A,B\}=AB+BA$
given in eq.~\eqref{PauliArithmetic}.
}
our result coincides 
with the corresponding four-quark operator of NRQCD calculated 
in \cite{Gerlach:2019kfo}.
In the latter calculation, $d_F^{abc}d_F^{abc}/N=(N^2-1)(N^2-4)/N^2$ is rewritten in terms of $C_A$, $C_F$ and $N$.
The results should agree, since only the HH region contributes in our calculation.
It would be useful to retain the color factor $d_F^{abc}d_F^{abc}/N$ as we give above, because for a general
gauge group it is an independent color factor, and also in some cases 
this color factor is related to a different physical origin from the other color factors.
(See, e.g., \cite{Moch:2004pa,Anzai:2010td}.)
In the quark-antiquark scattering amplitude, $d_F^{abc}d_F^{abc}/N$ resides in the diagrams with 3-gluon
intermediate states; see Fig.~\ref{Fig:Diag-3gluonInterMed}.

Combined with the two-loop Hamiltonian in the non-annihilation channel calculated recently, the full two-loop quarkonium Hamiltonian is obtained.
For instance, it can be used for calculations of the hyperfine and fine splittings 
at order $\alpha_s^6 m$.
To avoid any confusion,
we clarify the convention of the spinor basis.
The Hamiltonian in the non-annihilation channel before expansion in $\epsilon$ is given in the form
\bea
H_{\rm NAn}=\frac{C_F \,\bar{\mu}^{-2\epsilon}}{ k^2} \sum_{i=1}^6 \sum_{j=1}^3
\sum_{n,\ell\ge 0}\left(g_R^2 \,\bar{\mu}^{2\epsilon}\right)^j W^{\rm NAn}_{\{i,j,n,2\ell\}} \left(\frac{k}{m}\right)^n 
\frac{(p^2)^\ell+(p'^2)^\ell}{2\,m^{2\ell}}\, \Lambda^{\rm NAn}_i\,.
\label{defW}
\eea
The spinor basis is defined in dimensional regularization as
\bea
&&
\Lambda^{\rm NAn}_1=\mathbb{I}\otimes\mathbb{I}\,,
 \nonumber\\ &&
\Lambda^{\rm NAn}_2=\sigma^a\,\sigma^b \otimes \sigma^a\,\sigma^b\,,
 \nonumber \\ &&
\Lambda^{\rm NAn}_3=\sigma^a\,\sigma^b\, \sigma^c\,\sigma^d \otimes \sigma^a\,\sigma^b\, \sigma^c\,\sigma^d
\,,
 \\ &&
\Lambda^{\rm NAn}_4=\frac{1}{m^2}\left(\vec{\sigma}\!\cdot\!\vec{k}\,\sigma^a \otimes \vec{\sigma}\!\cdot\!\vec{k}\,\sigma^a  \right)
 \,,
 \nonumber \\ &&
\Lambda^{\rm NAn}_5=
\frac{1}{m^2}\left(\vec{\sigma}\!\cdot\!\vec{p}^{\,\prime}\,\vec{\sigma}\!\cdot\!\vec{p} \otimes \mathbb{I}
+ \mathbb{I}\otimes \vec{\sigma}\!\cdot\!\vec{p}^{\,\prime}\,\vec{\sigma}\!\cdot\!\vec{p}\right)
\,,
\nonumber \\ &&
\Lambda^{\rm NAn}_6=\frac{1}{m^4}\left(\vec{\sigma}\!\cdot\!\vec{p}^{\,\prime}\,\vec{\sigma}\!\cdot\!\vec{p} \otimes  \vec{\sigma}\!\cdot\!\vec{p}^{\,\prime}\,\vec{\sigma}\!\cdot\!\vec{p}\right) 
\,.
\nonumber
\eea
Contraction with the spinors is defined as\footnote{ 
The right-hand side of eq.~\eqref{Eq:ContractionNAn} can also be
written as 
$\left(\xi'^\dagger \sigma^a\,\sigma^b\,  \xi \right) 
\left(\tilde{\zeta}^\dagger \sigma^b\,\sigma^a\,  \tilde{\zeta}' \right) 
$ in terms of the $\mathbf{2}^*$ representation of the antiquark spinors.
}
\bea
\bra{\xi',\zeta'}\bar{\Lambda}^{\rm NAn}_2\ket{\xi,\zeta}=
\left(\xi'^\dagger \sigma^a\,\sigma^b\,  \xi \right) 
\left({\zeta^{\prime}}^\dagger \sigma^a\,\sigma^b\,  \zeta \right) \,, 
~~~\text{etc}.
\label{Eq:ContractionNAn}
\eea
The Hamiltonian in the non-annihilation channel after expansion in $\epsilon$ is obtained by
setting $H_{\rm An} \to H_{\rm NAn}$ and $C^{\rm An} \to C^{\rm NAn}$ in
eq.~\eqref{Eq: HamiltonianAn} with the same $O_i$'s.
The Wilson coefficients, 
$W^{\rm NAn}_{\{i,j,n,2\ell\}}$ and $C^{\rm NAn}_{\{i,j,n,2\ell\}}$, are given in the Supplementary 
Material of \cite{Mishima:2024afk}.\footnote{
We note that $m^{\epsilon}$ is missing in the machine-readable expressions
in the accompanying file of \cite{Mishima:2024afk}.
Namely, we should set the hard master integrals as $i I_H \to m^{-2\epsilon} i I_H$ and 
$I_{HH}^i \to m^{-4\epsilon} I_{HH}^i$
to restore the correct mass dimensions.
The authors thank C.~Peset and A.~Maier for pointing this out.
}

\section*{Appendix: Parameters in two-loop Wilson coefficients}

The list of two-loop Wilson coefficients in the annihilation channel $W^{\rm An}_{\{i,3,n,2\ell\}}$,
which is given as a separate file \cite{MathematicaFile}, includes the following parameters,
in addition to those already explained in the text.
The one-loop counter term for the gauge coupling constant reads
\bea
&& 
\delta _1 Z_g=
\frac{2 \left(n_h+n_l\right)-11 C_A}{96 \pi^2 \epsilon }
\,. 
\eea
The two-loop master integrals are given by\footnote{
Essentially these are calculated in \cite{Piclum:2007an,Gerlach:2019kfo,Mishima:2024afk}.
Note, however, sign and normalization conventions vary.
}
\bea
&&
I_{{HH}}^a = \frac{e^{-2 \gamma_E  \epsilon }}{ (4 \pi )^{4-2 \epsilon }} 
\Biggl[
 \frac{3}{2 \epsilon ^2}+\frac{17}{4 \epsilon }+\frac{\pi
   ^2}{4}+\frac{59}{8}
   +\epsilon\left(-\zeta (3)+\frac{65}{16}+\frac{49 \pi
   ^2}{24}\right) 
\nonumber\\&&~~~~~~~~~~~~~~~~~~~~
   +\epsilon ^2
   \left(\frac{151 \zeta (3)}{6}-\frac{1117}{32}+\frac{475 \pi ^2}{48}+\frac{7 \pi
   ^4}{240}-8 \pi ^2 \log (2)\right)
\nonumber\\&&~~~~~~~~~~~~~~~~~~~~
+\epsilon
   ^3 \left(192 \text{Li}_4\left(\frac{1}{2}\right)+\frac{2125 \zeta (3)}{12}-\frac{\pi ^2
   \zeta (3)}{6}-\frac{3 \zeta (5)}{5}-\frac{13783}{64}
\right.
\nonumber\\&&~~~~~~~~~~~~~~~~~~~~ ~~~~~
\left.
   +\frac{3745 \pi ^2}{96}-\frac{103 \pi
   ^4}{96}+8 \log ^4(2)+16 \pi ^2 \log ^2(2)-52 \pi ^2 \log (2)\right)
   \Biggr]
\nonumber\\&&~~~~~~~~~~~~~~~~~~~~
+ {\cal O}(\epsilon^4)
\,,
\\&&
I_{{HH}}^b = \frac{e^{-2 \gamma_E  \epsilon }}{ (4 \pi )^{4-2 \epsilon }} 
\Biggl[
\frac{1}{\epsilon ^2}+\frac{2}{\epsilon }+\frac{11 \pi ^2}{12}-\frac{1}{2}
   +\epsilon 
   \left(\frac{181 \zeta (3)}{12}-\frac{85}{4}+\frac{17 \pi ^2}{24}+\frac{3}{2} \pi ^2 \log
   (2)\right)
\nonumber\\&&~~~~~~~~~~~~~~~~~~~~
   +\epsilon ^2 \left(-36 \text{Li}_4\left(\frac{1}{2}\right)+\frac{157 \zeta
   (3)}{24}-\frac{907}{8}-\frac{373 \pi ^2}{48}+\frac{167 \pi ^4}{72}
\right.
\nonumber\\&&~~~~~~~~~~~~~~~~~~~~ ~~~~~
\left.
   -\frac{3 \log
   ^4(2)}{2}+3 \pi ^2 \log ^2(2)+\frac{3}{4} \pi ^2 \log (2)\right)
\nonumber\\&&~~~~~~~~~~~~~~~~~~~~
+\epsilon ^3 \left(-18 \text{Li}_4\left(\frac{1}{2}\right)+72
   \text{Li}_5\left(\frac{1}{2}\right)-\frac{7733 \zeta (3)}{48}+\frac{2845 \pi ^2 \zeta
   (3)}{72}
\right.
\nonumber\\&&~~~~~~~~~~~~~~~~~~~~ ~~~~~
\left.
   +\frac{15329 \zeta (5)}{40}-\frac{7273}{16}-\frac{4975 \pi ^2}{96}+\frac{107 \pi
   ^4}{90}-\frac{3 \log ^5(2)}{5} -\frac{3 \log ^4(2)}{4}
\right.
\nonumber\\&&~~~~~~~~~~~~~~~~~~~~ ~~~~~
\left.
  +2 \pi ^2 \log ^3(2)+\frac{3}{2} \pi
   ^2 \log ^2(2)+\left(\frac{23 \pi ^4}{5}-\frac{123 \pi ^2}{8}\right) \log
   (2)\right)
   \Biggr]
+ {\cal O}(\epsilon^4)
\,,
\nonumber\\&&~~~~~~~~~~~~~~~~~~~~
\\&&
I_{{HH}}^c = 
-\frac{(4 \pi )^{2 \epsilon -4} \Gamma (3-4 \epsilon ) \Gamma (1-\epsilon )^2 \Gamma (\epsilon ) \Gamma (2 \epsilon -1)}{\Gamma (3-3 \epsilon ) \Gamma (2-2 \epsilon )}
\,,
\\&&
I_{{HH}}^d = 
\frac{e^{2 i \pi  \epsilon } \pi ^{2 \epsilon -4} \Gamma (1-\epsilon )^3 \Gamma (2 \epsilon -1)}{64 \Gamma (3-3 \epsilon )}
\,,
\\&&
I_{{HH}}^e = 
\frac{e^{-2 \gamma_E  \epsilon }}{ (4 \pi )^{4-2 \epsilon }}  \left[
\frac{3 \zeta (3)}{2}-\frac{i \pi ^3}{6}-\frac{2}{3} \pi ^2 \log (2)\right]
+ {\cal O}(\epsilon)
\,,
\\&&
I_{{HH}}^f = 
 -\frac{2^{4 \epsilon -9} e^{i \pi  \epsilon } \pi ^{2 \epsilon -\frac{5}{2}} \csc (\pi  \epsilon ) \Gamma (\epsilon -1)}{\Gamma \left(\frac{3}{2}-\epsilon \right)}
\,,
\eea
\bea
&&
I_{{HH}}^g = 
 \frac{e^{-2 \gamma_E  \epsilon }}{ (4 \pi )^{4-2 \epsilon }}  \Biggl[
 \frac{1}{2 \epsilon ^2}+\frac{1}{\epsilon }\left( \frac{5}{2}+i \pi -2 \log (2) \right)
\nonumber\\&&~~~~~~~~~~~~~~~~~~~~ ~~
 -\frac{13 \pi ^2}{12}+4 i \pi +\frac{19}{2}+4 \log ^2(2)-4 i \pi  \log (2)-8 \log (2)
\nonumber\\&&~~~~~~~~~~~~~~~~~~~~ ~~
+ \epsilon  \left(-\frac{77 \zeta (3)}{6}+\frac{65}{2}+12 i \pi -\frac{47 \pi ^2}{12}-\frac{i \pi ^3}{3}-\frac{16 \log ^3(2)}{3}+16 \log ^2(2)
\right.
\nonumber\\&&~~~~~~~~~~~~~~~~~~~~ ~~~~
\left.
+8 i \pi  \log ^2(2) -24 \log (2)-16 i \pi  \log (2)+\frac{13}{3} \pi ^2 \log (2)\right)
 \Biggr]
\nonumber\\&&~~~~~~~~~~~~~~~~~~~~ ~~~~
+ {\cal O}(\epsilon^2)
\,,
\\&&
I_{{HH}}^h = 
\frac{e^{-2 \gamma_E  \epsilon }}{ (4 \pi )^{4-2 \epsilon }}  \left[\frac{21 \zeta (3)}{8}-\frac{i \pi ^3}{8}-\frac{1}{2} \pi ^2 \log (2)\right]
+ {\cal O}(\epsilon)
\,,
\\&&
I_{{HH}}^i = 
\frac{2^{4 \epsilon -10} e^{2 i \pi  \epsilon } \pi ^{2 \epsilon -1} \csc ^2(\pi  \epsilon )}{\Gamma \left(\frac{3}{2}-\epsilon \right)^2}
\,.
\eea

\section*{Acknowledgement}
The authors are grateful to G.~Mishima for discussion. 
One of the authors (YS) is also grateful to A.~Pineda and A.~Maier for useful comments.
All the figures were created with \texttt{TikZ-FeynHand}~\cite{Ellis:2016jkw,Dohse:2018vqo}.
This work was supported in part
by JSPS KAKENHI Grant Numbers 
JP22K03604, JP23K03404 and JP24K07055.

\end{document}